# THE NATIONAL IGNITION FACILITY: STATUS AND PLANS FOR LASER FUSION AND HIGH-ENERGY-DENSITY EXPERIMENTAL STUDIES

E.I. Moses
LLNL, Livermore, CA 94550, USA

Abstract

The National Ignition Facility (NIF) currently under construction at the University of California Lawrence Livermore National Laboratory (LLNL) is a 192-beam, 1.8-megajoule, 500-terawatt, 351-nm laser for inertial confinement fusion (ICF) and high-energy-density experimental studies. NIF is being built by the Department of Energy and the National Nuclear Security Agency (NNSA) to provide an experimental test bed for the U.S. Stockpile Stewardship Program to ensure the country's nuclear deterrent without underground nuclear testing. The experimental program will encompass a wide range of physical phenomena from fusion energy production to materials science. Of the roughly 700 shots available per year, about 10% will be dedicated to basic science research. Laser hardware is modularized into line replaceable units (LRUs) such as deformable mirrors, amplifiers, and multi-function sensor packages that are operated by a distributed computer control system of nearly 60,000 control points. The supervisory control room presents facility-wide status and orchestrates experiments using operating parameters predicted by physics models. A network of several hundred front-end processors (FEPs) implements device control. The object-oriented software system is implemented in the Ada and Java languages and emphasizes CORBA distribution of reusable software objects. NIF is currently scheduled to provide first light in 2004 and will be completed in 2008.

## 1 INTRODUCTION

The NIF currently under construction at LLNL will be a U.S. Department of Energy and NNSA national center to study inertial confinement fusion and the physics of extreme energy densities and pressures. It will be a vital element of the NNSA Stockpile Stewardship Program (SSP), which ensures the reliability and safety of U.S. nuclear weapons without full-scale underground nuclear testing. The SSP will achieve this through a combination of above-ground test facilities and powerful computer simulations using the NNSA's Accelerated Scientific Computing Initiative (ASCI). In NIF, up to 192 extremely powerful laser beams will compress small fusion targets to conditions in which they will ignite and burn, liberating more energy than is required to initiate the fusion reactions. NIF experiments will allow the study of physical processes at temperatures approaching 100 million K and 100 billion times atmospheric pressure. These conditions exist naturally only in the interior of stars and in nuclear weapons explosions.

## 2 DESCRIPTION OF NIF

The NIF is shown schematically in Figure 1. NIF consists of four main elements: a laser system and optical components; the target chamber and its experimental systems; an environmentally controlled building housing the laser system and target area; and an integrated computer control system.

NIF's laser system features 192 high-power laser beams. Together, the laser beams will produce 1.8 million joules (approximately 500 trillion watts of power for 3 nanoseconds) of laser energy in the near-ultraviolet (351 nanometer wavelength). Currently the largest operating laser is the Omega Laser at the University of Rochester's Laboratory for Laser Energetics. Omega consists of 60 laser beams delivering a total of 40 kilojoules of energy. Figure 2 shows one of the 192 laser beams, detailing the key technologies that make NIF possible. A NIF laser beam begins with a very modest nanojoule energy pulse from the master oscillator, a diode-pumped fiber laser system that can provide a variety of pulse shapes suitable for a wide range of experiments, from ICF implosions to high-energy extended pulses for weapons effects experiments. The master oscillator pulse is shaped in time and smoothed in intensity and then transported to preamplifier modules (PAMs) for amplification and beam shaping. Each PAM first amplifies the pulse by a factor of one million (to a millijoule) and then boosts the pulse once again, this time to a maximum of 22 joules, by passing the beam four times through a flashlamp-pumped amplifier. There are total of 48 PAMs on NIF, each feeding a "quad" of four laser beams.

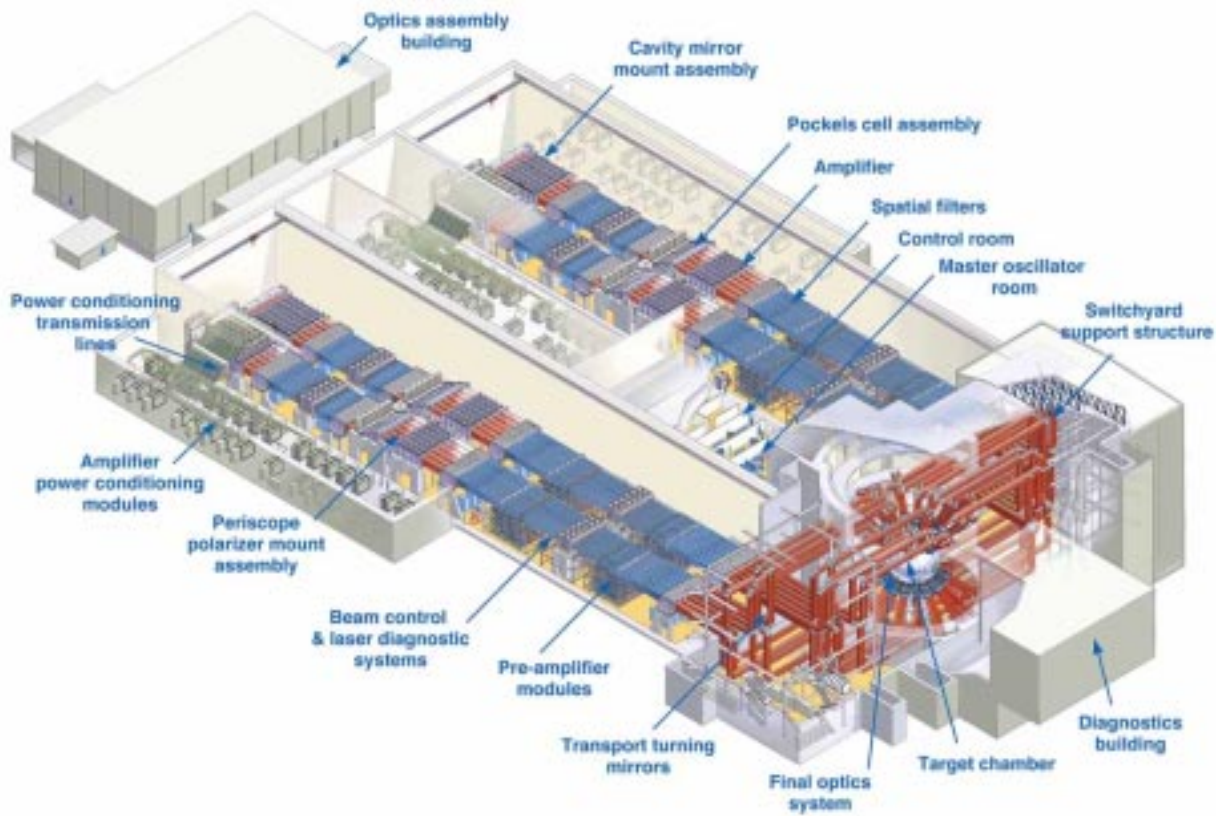

Figure 1: Schematic view of the National Ignition Facility showing the main elements of the laser system. The 10-meter diameter target chamber on the right side of the illustration sets the scale for the facility.

From the PAM, the laser beam next enters the main laser system, which consists of two large amplifier units—the power amplifier and the main amplifier. These amplifier systems are designed to efficiently amplify the nominal 1-joule input pulse from the PAM to the required power and energy, maintaining the input beam's spatial, spectral, and temporal characteristics. The amplifiers, with 16 glass slabs per beam, are arranged with 11 slabs in the main amplifier section and 5 slabs in the power amplifier section. Together these amplifiers provide 99.9% of NIF's energy. The amplifiers use 42-kilogram slabs, 46 cm × 81 cm × 3.4 cm, of neodymium-doped phosphate glass set vertically on edge at Brewster's angle to minimize reflective losses in the laser beam. The slabs are stacked four high and two wide to accommodate a "bundle" of eight laser beams (Figure 3).

The slabs are surrounded by vertical arrays of flashlamps measuring 180 cm in length. NIF's 192 laser beams require 7600 flashlamps and 3072 glass slabs. Each flashlamp is driven by 30,000 joules of electrical energy. The intense white light from the flashlamps excites the neodymium in the laser slabs to provide optical gain at the primary infrared wavelength of the laser. Some of the energy stored in the neodymium is released when the laser beam passes through the slab. The flashlamps and amplifier slabs will be cooled between shots using nitrogen gas. NIF will be able to shoot once every 8 hours; however, a shot rate enhancement program funded by collaborators from the United Kingdom is working to increase this rate to once every four hours.

The NIF amplifiers receive their power from the Power Conditioning System (PCS), which consists of the highest energy array of electrical capacitors ever assembled. The system's design is the result of collaboration between Sandia National Laboratories in Albuquerque, LLNL, and industry. The PCS will occupy four capacitor bays (Figure 1) adjacent to the laser bays. Each PCS module has eight 20-capacitor modules, delivering 1.7 megajoules per module, which power the flashlamps for one beam. The system must deliver over 300 megajoules of electrical energy to the flashlamp assemblies in each laser beam. Recent tests on a prototype PCS and flashlamp system have been fired over 7000 times at a rate of 1200 shots per month.

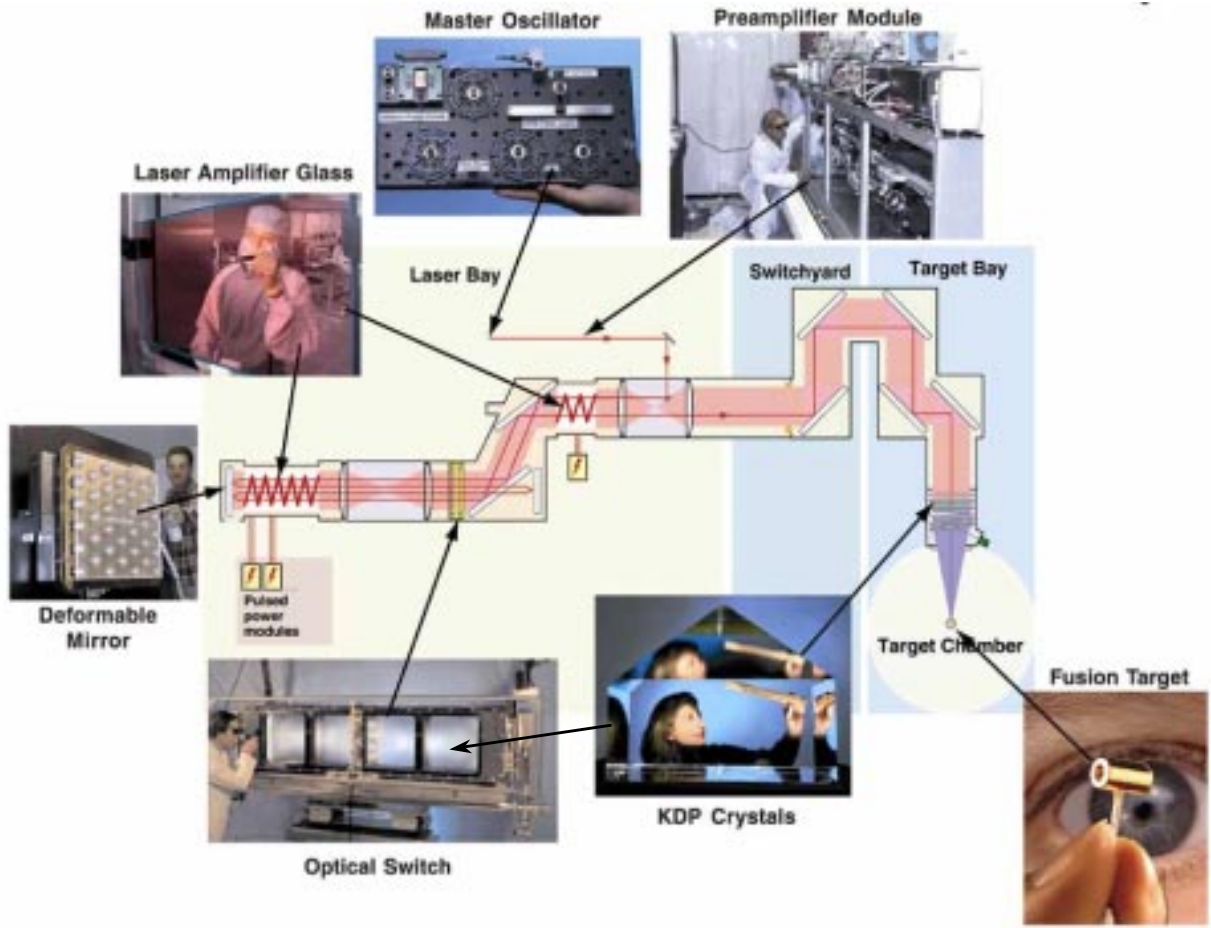

Figure 2: Schematic representation of a NIF laser beam line highlighting some of the key technology developments.

A key component in the laser chain is a kind of optical switch called a plasma electrode Pockels cell (PEPC), which allows the beam to pass four times through the main amplifier cavity. This device uses electrically induced changes in the refractive index of an electro-optic crystal, made of potassium dihydrogen phosphate (KDP). When combined with a polarizer, the PEPC allows light to pass through or reflect off the polarizer. The PEPC will essentially trap the laser light between two mirrors as it makes two round-trip passes through the main amplifier system before being switched out to continue its way to the target chamber. The PEPC consists of thin KDP plates sandwiched between two gas-discharge plasmas that are so tenuous that they have no effect on the laser beam passing through the cell. Nonetheless, the plasmas serve as conducting electrodes, allowing the entire surface of the thin crystal plate to charge electrically in about 100 nanoseconds so the beam can be switched efficiently. Figure 2 shows a prototype 4-cell PEPC (optical switch) that will be stacked vertically in a single unit called a line-replaceable unit (LRU).

All major laser components are assembled in clean modules called LRUs. These LRUs contain laser optics, mirrors, lenses, and hardware such as pinhole filter assemblies. All LRUs are designed to be assembled and installed into NIF's beampath infrastructure system, the exoskeleton of NIF, while retaining the high level of cleanliness required for proper laser operation. LLNL's industrial partner, Jacobs Facilities, Inc. is responsible for the installation, integration, and commissioning of the NIF laser beampath infrastructure in a way that ensures that the required cleanliness levels are maintained throughout the installation and commissioning phase of the Project.

The NIF target area consists of the 10-meter-diameter high-vacuum target chamber shown in Figure 4. The target chamber features a large number of laser entry ports as well as over 100 ports for diagnostic instrumentation and target insertion. Each laser entry port allows a quad of four laser beams to be focused to the center of the target chamber through a final optics assembly (FOA).

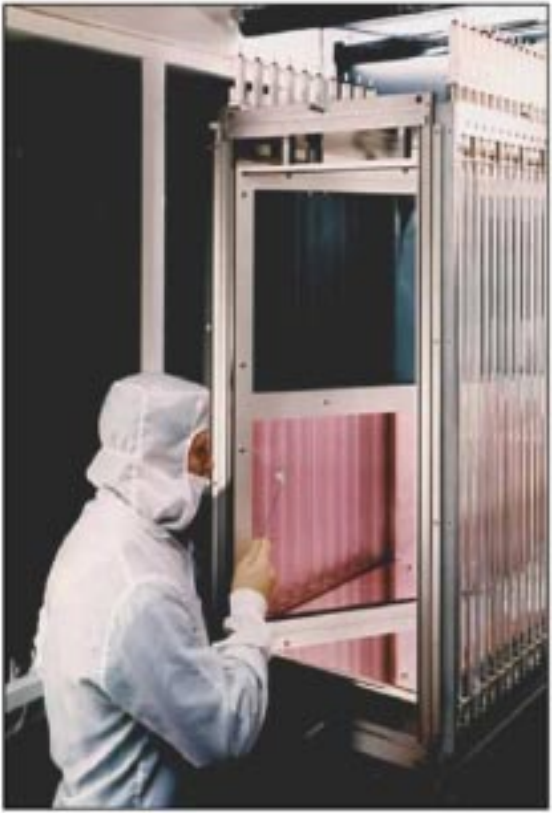 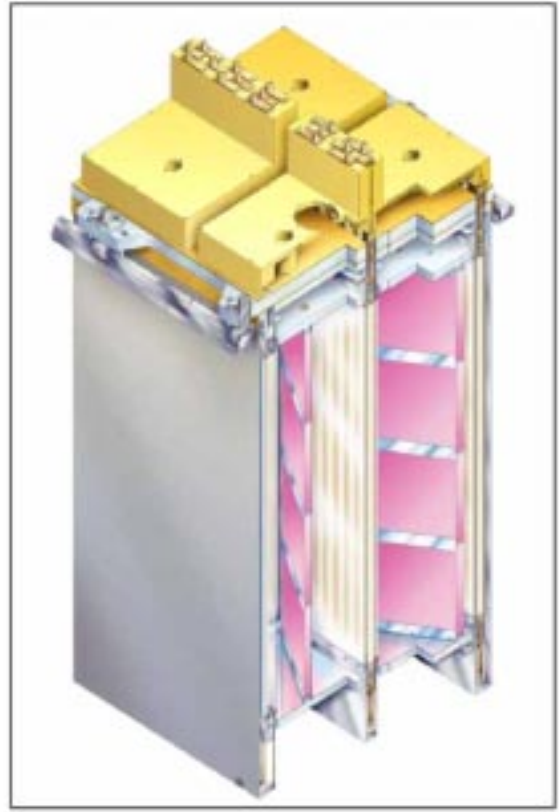

Figure 3. The photograph on the left shows an amplifier used on Beamlet, the scientific prototype of NIF. The illustration on the right shows the NIF 2 × 4 amplifier in cutaway view.

The FOA is a precision optical assembly containing beam smoothing gratings, additional KDP and deuterated KDP plates for second- and third-harmonic generation to convert the infrared laser light into the ultraviolet, the final focus lens, debris shields, and a vacuum gate valve for each beam. The NIF target chamber and final focusing system has been designed with maximum flexibility for experimental users. During initial operation, NIF is configured to operate in the "indirect-drive" configuration, which directs the laser beams into two cones in each of the upper and lower hemispheres of the target chamber. This configuration is optimized for illuminating a fusion capsule mounted inside a cylindrical hohlraum and using x-rays generated from the hot walls of the hohlraum to implode the capsule indirectly. NIF can also be configured in a "direct-drive" arrangement of beams by moving some quads of beams from the upper and lower beam cones into a more symmetric arrangement of beams. Direct-drive ignition requires better energy and power balance between laser beams and better beam smoothing and focusing, but the simpler geometry makes direct-drive inertial confinement fusion more attractive for ultimately producing a viable power production plant.

## 3 NIF CONTROL SYSTEMS

The Integrated Computer Control System (ICCS) for the NIF is a layered architecture of 300 FEP coordinated by supervisor subsystems. FEP computers incorporate either VxWorks on PowerPC or Solaris on UltraSPARC processors that interface to over 45,000 control points attached to VME-bus or PCI-bus crates respectively. Supervisory computers use Solaris workstations to implement coordination, database services, and user interfaces. Typical devices are stepping motors, transient digitizers, calorimeters, and photodiodes. The front-end implements an additional segment comprised of 14,000 control points for industrial controls including vacuum, argon, synthetic air, and safety interlocks using Allen-Bradley programmable logic controllers. The computer network uses Ethernet for control and status signals and is augmented with asynchronous transfer mode to deliver video streams from 500 sensor cameras within the laser to operator workstations. Software uses CORBA distribution to define a framework that incorporates services for archiving, machine configuration, graphical user interface, monitoring, event logging, scripting, alert management, and access control. Software coding uses a mixed language environment of object-oriented Ada95 and Java. The code is one-third complete at over 300 thousand source lines.

## 4 NIF PROJECT STATUS

NIF is currently over four years into its construction. The conventional building construction is nearly complete. The attached 8000-square-foot Class-100 clean room Optics Assembly Building is undergoing commissioning of LRU assembly, handling, and transport equipment. Both large laser bays are operating under Class-100,000 clean room protocols. Over 1500 tons of beampath infrastructure have been installed in the laser bays. The NIF Project is entering the installation and commissioning phase. First light, which is defined as the first quad of four laser beams focused to target chamber center, is scheduled for June 2004. Full completion of all 192 laser beams is scheduled for September 2008. In the time between first light and project completion, approximately 1500 experiments in support of the SSP, inertial confinement fusion, high-energy-density physics, weapons effects, inertial fusion energy, and basic science will have been performed.

After project completion, NIF is expected to provide approximately 750 shots per year for a wide variety of experimental users. Recently, NIF was designated as a National User Facility with the support of the NNSA Office of Defense Programs. A National User Support Organization is being put in place to provide the necessary interface between the user communities and the national NIF Program. The first Director of NIF is Dr. George H. Miller, from LLNL, who also serves as the Associate Director for NIF Programs at LLNL.

## 5 CONCLUSIONS

The National Ignition Facility has come a long way since the first DOE critical decision in January 1993 affirmed the need for NIF and authorized the conceptual design process. In that time, NIF has met every scientific and technical challenge and is now in the final stages of design and construction prior to commencing installation of the 192 laser beams. By 2004 this unique facility will be providing the first glimpses under repeatable and well-characterized laboratory conditions of phenomena heretofore only found in the most extreme environments imaginable.

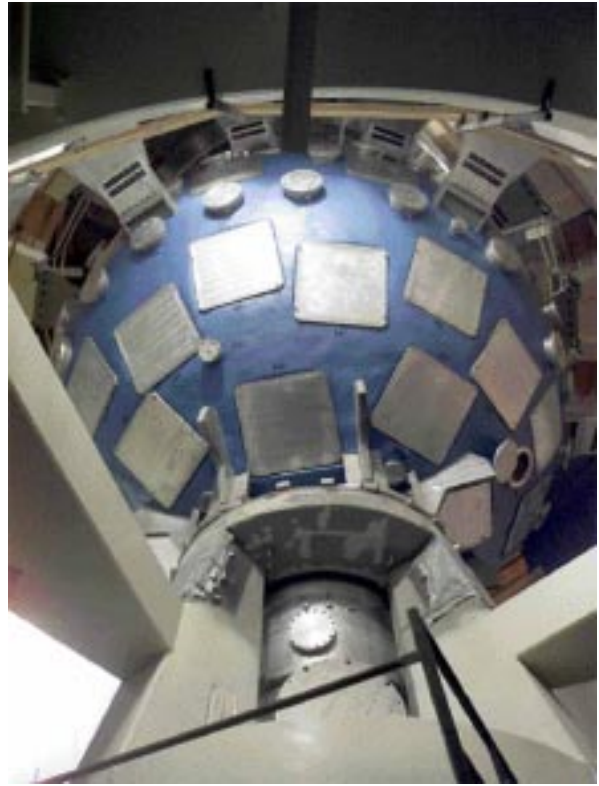

Figure 4: NIF's 10-meter-diameter target chamber mounted in the target bay and viewed from below.

## 6 ACKNOWLEDGEMENTS